\documentclass[preprint]{revtex4-1}

\usepackage{graphicx}
\usepackage{dcolumn}
\usepackage{bm}

\usepackage[utf8]{inputenc}
\usepackage[T1]{fontenc}

\usepackage{etoolbox}
\usepackage{amsmath}
\usepackage{subcaption}
\usepackage{tabularx}

\makeatletter
\def\@email#1#2{%
 \endgroup
 \patchcmd{\titleblock@produce}
  {\frontmatter@RRAPformat}
  {\frontmatter@RRAPformat{\produce@RRAP{*#1\href{mailto:#2}{#2}}}\frontmatter@RRAPformat}
  {}{}
}%
\makeatother

\usepackage[colorlinks=true, linkcolor=blue, citecolor=blue, urlcolor=blue]{hyperref}

\usepackage{newtxtext,newtxmath}
\usepackage{comment}
\usepackage{float}
\setcitestyle{super}

\begin{document}

\title[Sample title]{Turbulent Flows in Electron Hydrodynamics: Conductivity and Vorticity}

\author{Kanad Bhattacharya}
\affiliation{Deeksha STEM School, Bengaluru, India}
\email{80051251094\_kanad@deekshastemschool.edu.in}

\date{November 2025}

\begin{abstract}
In this article, we attempt to understand various aspects of turbulent flows in electron hydrodynamics. We analyze a rectangular channel geometry in the presence of an electric field and a Corbino geometry in the presence of a magnetic field. In the former geometry, we analyze the conductivity of the fluid as well as the frequency spectrum of perturbations about the Poiseuille flow. While the normal Poiseuille flow has an associated conductivity which scales as $W^2$, we find a correction which scales as $W^4$ in the case of non-linear flows, where $W$ is the characteristic length of the system. In the Corbino geometry, we analyze the velocity, vorticity and magnetic fields. We find that the vorticity can span across a wide range near the edge of the geometry, a behavior that can be reflected in the velocity and magnetic fields.
\end{abstract}

\maketitle

\section{Introduction}
\label{sec:int}

The concept of using hydrodynamics to describe the motion of electrons was first suggested by Gurzhi in 1963. \cite{Gu963} The hydrodynamic regime works well when the momentum conserving electron-electron interactions dominate over the momentum relaxing electron-impurity and electron-phonon interactions. More specifically, it corresponds to the viscous scenario when $l_{\mathrm {MC}}\ll W \ll l_{\mathrm {MR}}$ as compared to the Ohmic regime $l_{\mathrm {MR}},l_{\mathrm {MC}}\ll W$ and the ballistic regime when $W \ll l_{\mathrm {MR}}, l_{\mathrm {MC}}$, where $l_{\mathrm {MC}}$ and $l_{\mathrm {MR}}$ denote the length scales associated with momentum conserving and momentum relaxing interactions while $W$ represents the characteristic length of the system. \cite{GMK18} Gurzhi predicted that resistivity can decrease with temperature, which was experimentally proven. \cite{MD994} In recent years, there has been a surge of interest in this field due to advancements in experiments. For instance, viscous flows have been obtained in graphene and $\mathrm{PdCoO}_2$ samples. \cite{BTK16,MKN16}  Various interesting phenomenon such as negative local resistance and the violation of the Wiedemann--Franz law have been observed. \cite{BTK16,CSW16}

Most theoretical approaches are centered around various variants of the Navier--Stokes equations. A variety of geometries and boundary conditions have been analyzed, where conductivity of the electron fluids is widely studied. \cite{TTGP15,PTG15,TVP14, SNS17} Some works examine more general viscosity tensors and conditions for the presence of hydrodynamic behavior \cite{VJA20,EDD24} while others explore the other thermoelectric coefficients related to transport. \cite{AKS11,LLA20,LLA22a,LLA22b} Changing the quantum mechanical settings under which the flow takes place also leads to interesting phenomena. \cite{HEI21} Many exotic situations are analyzed using the Boltzmann equation, which is assumed to be exact in most situations. \cite{CL19,NL24}

In this article, we analyze the effect of non-linear flows. Understanding turbulence in the context of electron hydrodynamics is one of the upcoming challenges of the field. \cite{FS24}  We stress that developing a theory that parallels Kolmogorov and Kraichnan's theories of classical turbulence will involve classical, statistical and quantum mechanical ideas as reflected by the following inequality \cite{KSS05}

\begin{equation}
    \frac{\mu}{s}\ge \frac{\hbar}{4\pi k_B},
\end{equation}
where $\mu$ is the shear viscosity, $s$ is the entropy volume density.

Among the literature on non-linear flows \cite{DS993,MPL21,FGS22,CCJ21,GPS18,EKS24} there seems to be an interest in the Dyakonov-Shur instability. The main purpose of such a phenomenon is the production of Terahertz electromagnetic radiation using plasma oscillations. Terahertz radiation can have immense applications in industry. Typically, the geometry analyzed is rectangular with boundary conditions of the form \cite{MPL21}
\begin{equation*}
    \begin{aligned}
n|_{x=0} &= n_{0}, \\
nu|_{x=L} &= n_{0} u_{0}, \\
\partial_{x} (nu)|_{x=0} &= 0 .
\end{aligned}
\end{equation*}
where $n$ is the number density, $L$ is the length of the system and $u$ is the $x$ component of the velocity field. Sometimes, such approaches involve assumptions such as $\partial_y \mathbf u=\mathbf 0$ or $\partial_\theta \mathbf u=\mathbf 0$. However, if we try removing such symmetries in order to understand the more nuanced effects of turbulent flows, the equations can become analytically intractable.

In this work, we try to circumvent such issues by considering two separate geometries discussed in sections \ref{sec:rec} and \ref{sec:crb} to understand various aspects of turbulent flows. At the heart of our work are the Navier--Stokes equations
\begin{subequations}
    \begin{eqnarray}
        \partial_t n+\nabla_{\mathbf r} (n\mathbf u)=0,\\
          \partial _t \mathbf u+ \mathbf u \cdot \nabla_{\mathbf r} \mathbf u=\nu \nabla_{\mathbf r}^2 \mathbf{u}+\mathbf F,
    \end{eqnarray}
    \label{eq:NS}
\end{subequations}
where $n$ is the 2D number density, $\mathbf u$ is the velocity field, $\nu$ is the kinematic viscosity and $\mathbf F$ is the external force per mass.
We first consider a rectangular geometry in the presence of an electric field. We understand how the conductivity scales with the size of the system using approximations of the Navier--Stokes equations and (later) using Kraichnan's theory \cite{Kr67} of 2D turbulence alongside the Boltzmann equation
\begin{equation}
    \partial_t f+\mathbf v \cdot \nabla_{\mathbf r} f+ \frac{\bar {\mathbf F}}{\hbar} \cdot \nabla_{\mathbf k} f=\hat{C}[f],
    \label{eq:BTE}
\end{equation}
where $f$ is the Boltzmann distribution function, $\bar{\mathbf F}$ is the external force and $\hat C$ is the collision operator.
We also follow related works \cite{MPL21,FGS22} and analyze the frequency of perturbations about the equilibrium flow (the Poiseuille flow). The second geometry that we analyze is the Corbino geometry under the presence of a magnetic field, where linearized approaches can be addressed analytically if one works with the vorticity field. However, the velocity and magnetic fields due to the motion of electrons need to be determined numerically.

\section{Conductivity and Perturbation Frequencies in a Rectangular Geometry}
\label{sec:rec}

In this section, we analyze how the presence of an electric field affects turbulent flows. We consider an electron fluid of kinematic viscosity $\nu$ with electron effective mass $m$  in a rectangular channel of width $2W$ in the presence of an electric field $E \mathbf{\hat x}$. The Poiseuille flow is of central importance in our analysis, which occurs when the flow is in a steady state. The Navier--Stokes equations \eqref{eq:NS} in this case, reduce to 

\begin{equation}
    \nu \partial^2_y\mathbf u=\frac {eE\mathbf{\hat x}}{m}.
     \label{eq:Pod}
\end{equation}
while the velocity profile is given by \cite{SNS17}
\begin{equation}
    \mathbf u=\frac{eE}{2 m \nu}(y^2-W^2)\mathbf {\hat x}\equiv \bar u_x\mathbf{\hat x},
    \label{eq:Pos}
\end{equation}
under boundary conditions of the form $u_x|_{y=\pm W}=0$. Observe that the conductivity scales as $\sigma=\frac {-ne\langle v\rangle}{E} \propto W^2$ with the channel width. The following two sections analyze how turbulent flows affect this scaling of the conductivity with the system size. Meanwhile, in the third section, we understand how linearized approaches (employed in the previous sections) help us find the spectrum of perturbation frequencies around the Poiseuille flow. 

\subsection{Conductivity in Weak Turbulence}
\label{ssc:wtb}

In this subsection, we try to understand the contribution of the advective term. However, we make some simplifying assumptions: we set $\partial_t \mathbf u=0$, $\partial_x \mathbf u=0$ and (implicitly) relax the notion of incompressibility. As a first approximation, we introduce a perturbation of the velocity field $\mathbf u'$ about the Poiseuille flow. The linearized Navier--Stokes equations is given by
 \begin{equation}
     \bar u_x\partial_x \mathbf u' +u'_y\partial_y\bar u_x \mathbf{\hat x}+\nu \nabla^2 \mathbf u'=\mathbf 0.
     \label{eq:Pld}
 \end{equation}
 which reduces to

\begin{subequations}

\begin{eqnarray}
    \partial_y^2 u'_y=0, \label{eq:Ply}\\
    u'_y \partial_y \bar u_x+\partial_y^2u'_x=0. \label{eq:Plx}
\end{eqnarray}

\end{subequations}

Observe that \eqref{eq:Ply} implies $u'_y \propto y$, which along with \eqref{eq:Plx} implies $u'_x \propto y^4$. Hence, we obtain a correction to the Poiseuille conductivity of which scales as $W^4$.

As it turns out, this asymptotic obtained by such a crude calculation is reflected in more sophisticated calculations. For instance, instead of linearizing about the Poiseuille flow, we can directly solve the following equation exactly
\begin{equation}
    \mathbf u\cdot \nabla \mathbf u=\nu\nabla^2\mathbf u-\frac{eE\mathbf{\hat x}}{m},
    \label{eq:Ped}
\end{equation}
under the condition that $\partial_x \mathbf u=0$. 
We are interested in the following solution (which is physically only applicable in narrow channels)

\begin{subequations}
    \begin{eqnarray}
         u_y=\alpha \tan \left(\frac{\alpha y}{2\nu} \right),\\
          u_x=\frac{eE}{m\alpha}\left(y \tan\left(\frac{\alpha y}{2\nu}\right)- W\tan\left(\frac{\alpha W}{2\nu} \right) \right),
    \end{eqnarray}
    \label{eq:Pes}
\end{subequations}

where we have introduced a parameter $\alpha$ that needs to be determined through boundary conditions. Upon using a power series expansion, we see a fourth-order correction in $y$ to the Poiseuille flow. Hence, we once again recover a correction of the form $W^4$ to the conductivity (assuming $\alpha$ remains constant as we scale the system).

However, the issue with the solution \eqref{eq:Ped} is that it does not describe strongly turbulent flows due to the symmetries that we have assumed in our solutions. More sophisticated calculations involving variants of the Orr Sommerfeld equations can be done; however, these linearized approaches may prevent us from fully understanding the contribution of the advective term.  In the next section, we address this with calculations done using Kraichnan's theory of turbulence.

\subsection{Conductivity in Strong Turbulence}
\label{ssc:stb}
As stated in \cite{LF18} the hydrodynamics involved is a classical phenomenon in a quantum mechanical setting. However, the questions that classical hydrodynamics aims to answer are different from those of electron hydrodynamics. For instance, determining transport coefficients and electromagnetic fields is more relevant than understanding correlation functions and energy scaling laws. The challenge then becomes to incorporate theories of classical turbulence, such as Kraichnan's theory, to suit our needs (of computing the conductivity).

This is easier said than done. The notion of using dimension analysis to solve such a problem falls short when one realizes that there are too many relevant (length) scales: the Gurzhi length, the product of the Fermi velocity with some relevant timescale, the viscosity dissipation cutoff length, the length scale of the system, etc. The idea we propose here is to obtain a correction to the Boltzmann distribution function, which accounts for turbulent flows by using the energy spectrum from Kraichnan's theory, to obtain the correction to the conductivity. 

The energy spectrum per unit mass $E(k)$ associated with the inverse length scale $k$ is associated with two cascades: the enstrophy cascade as well as the energy cascade . For the enstrophy cascade we have $E(k)\sim \eta^{2/3}k^{-3}$, while for the energy cascade we have $E(k)\sim \varepsilon^{2/3}k^{-5/3}$. \cite{Kr67} Here $\eta$ is the rate of cascade of enstrophy per mass while $\varepsilon$ is the rate of cascade of energy per mass. The relevant scales for the cascades is given by $k_{\nu}>k_{\eta}>k_{f}>k_{\varepsilon}>k_{\alpha}$, where $k_{\nu}$ is the viscosity cutoff, $k_\eta$ is a scale corresponding to the enstrophy cascade, $k_f$ is the scale associated with external forces, $k_{\varepsilon}$ is a scale corresponding to the energy cascade and $k_{\alpha}$ is the scale associated with frictional damping. \cite{BE12} In the case where we are applying a uniform electric field, one would expect $k_f\sim 1/W$. Furthermore, if the damping effects can be ignored, one can argue that the enstrophy cascade is the dominant contribution.

Now, we obtain the Boltzmann distribution function from $E(k)$. We write 

\begin{equation}
    \int dk \ E(k)=\frac{1}{2}\langle{u}^2\rangle
    \label{eq:Eu^2},
\end{equation}
as well as \cite{Ka07}
\begin{equation}
    \int \frac{{d^2}\mathbf k}{4\pi^2} \mathbf \epsilon(\mathbf k) f(\mathbf k)=\int \frac{dk}{2 \pi}k \epsilon f =\frac{1}{2}nm \langle  u^2 \rangle.
    \label{eq:fu^2}
\end{equation}
Here $\epsilon(\mathbf k)$ is the wave number dependent dispersion energy (which we assume scales as $\epsilon\propto k^{\alpha}$ for some non-negative real number $\alpha$), $f(\mathbf k)$ is the (spatially averaged) Boltzmann function and $\langle \bullet \rangle$ denotes spatial average. Observe that if we compare integrands in \eqref{eq:Eu^2} and \eqref{eq:fu^2}, we arrive at the following expression for the distribution function

\begin{equation}
    f\approx A\frac{  nm \eta^{2/3}}{k^4\epsilon}+f_e,
    \label{eq:fAk}
\end{equation}
where $f_e$ is the distribution function associated with stabler flows and $A$ is a dimensionless constant. We shall be ignoring the contribution of $f_e$ in all our calculations since we are interested in $f$ to derive higher-order corrections. Observe that for our derivation of \eqref{eq:fAk} to hold good, the upper and lower bounds of the integrals in \eqref{eq:Eu^2} and \eqref{eq:fu^2} must be asymptotically similar. This can be justified as follows. Note that the lower bound of \eqref{eq:Eu^2} is $\sim 1/W$. \cite{LL987} Meanwhile, if we were to approximate the integral in \eqref{eq:fu^2} as a sum, it would range over $k_j=j\pi/W$, where $j$ is an integer. Hence, the lower bound in \eqref{eq:fu^2} can also be considered to be $\sim 1/W$. One can similarly argue that the upper bounds (which do not have $W$ dependence) must also be asymptotically similar and can be approximated as being infinite. \nocite{LKM21} \footnote{The asymptotic similarity of the bounds can be further motivated by the fact that the Gurzhi length \cite{LKM21} and the viscosity cutoff length  \cite{BE12} are proportional to $\nu^{1/2}$.} Hence, with \eqref{eq:fAk} established, we now turn our attention back to computing the conductivity. Let $f_0=A\frac{  nm \eta^{2/3}}{k^4\epsilon}$.

We follow the procedure given in \cite{GY19} for our calculation. The idea is to perturbatively write the Boltzmann distribution function under the influence of an electric field as $f=f_0-\phi\partial_{\epsilon}f_0$. Solving the Boltzmann equation \eqref{eq:BTE} we arrive at

\begin{equation}
    \phi=-e\tau E(\mathbf {\hat x}\cdot \mathbf v(\mathbf k))
    \label{eq:phi}
\end{equation}
where $\tau$ is a characteristic time associated with the collision term and $\mathbf v(\mathbf k)=\nabla_{\mathbf k}\epsilon/\hbar$ is the group velocity. Hence, we arrive at the conductivity 

\begin{equation}
    \sigma=\frac{J}{E}=-\frac{e}{E} \int \frac{ d^2 \mathbf k}{4\pi^2}(\mathbf v(\mathbf k)\cdot \hat{\mathbf{x}})f(\mathbf k)=\frac{e^2\tau}{2\hbar^2}\int \frac{ dk}{2\pi} k \left(-\frac{\partial f_0}{\partial \epsilon}\right)(\nabla_{\mathbf k}\epsilon)^2
    \label{eq:sJE}.
\end{equation}
Hence, we find the correction to the conductivity 

\begin{equation}
    \sigma\sim \frac{e^2\tau nm\eta^{2/3}W^4}{\hbar^2}
    \label{eq:seW}
\end{equation}
which once again displays the $W^4$ behavior (as long as $n \eta^{2/3}$ is constant). \footnote{It must be pointed out that turbulent flows may not always increase the conductivity, and that the correction of the form $W^4$ could be subtracted from the conductivity, instead of being added. For instance, in \eqref{eq:Pes} if one replaces $\tan$ with $\tanh$, the solution is still valid and the conductivity is reduced.}  Notice that such a derivation does not apply to macroscopic fluids, as the scale at which the Navier--Stokes equations describes a fluid is much greater than the microscopic scale at which the Boltzmann equation operates (which is not the case in electron fluids).

\subsection{Frequency of perturbations}
\label{ssc:frq}
Although the Orr-Sommerfeld equation may not be particularly helpful in analyzing conductivity, it can help us arrive at an analytic approximation for the frequency spectrum of perturbations. Previously the Orr-Sommerfeld equation has been used in stability analysis. \cite{MMS19} We consider a correction of the form $e^{i \beta x-i \Omega t} \varphi$ to the streaming function $\psi$ for the velocity field $\mathbf u=\nabla \times (\psi \mathbf{\hat z}) $. Linearizing, we obtain the Orr-Sommerfeld equations for the Poiseuille flow \cite{OS971}

\begin{equation}
    \frac{\nu}{i\beta }(\partial_y^2 -\beta^2)^2\varphi=\left(-\frac{eEW^2}{2m \nu}+\frac{eEy^2}{2m \nu}-\frac{\Omega}{\beta}\right)(\partial_y^2-\beta^2)\varphi-\frac{eE}{m \nu}\varphi.
    \label{eq:OrS}
    \end{equation}
If we focus around the center of our channel at $y=0$, we can ignore the $y^2$ term. We furthermore assume that $\varphi \propto e^{ij\pi y/W}$ where $j$ is an integer. Hence, solving for $\Omega$ in terms of $\beta$ and $j$, we arrive at 

\begin{equation}
    \Omega(j ,\beta)=-\frac{e\beta EW^2}{2m\nu}+\frac{\frac{e\beta E}{m\nu}-i\nu\left(\frac{j^2\pi^2}{W^2}+\beta^2\right)^2 }{\left(\frac{j^2\pi^2}{W^2}+\beta^2\right)}.
    \label{eq:Obn}
\end{equation}

\section{Vorticity in Corbino Geometry}
\label{sec:crb}

In this section, we turn our attention to the Corbino geometry with our focus centered around the vorticity field $\omega=(\nabla \times \mathbf u)\cdot \mathbf{\hat z}$ and the vorticity equation
\begin{equation}
    \partial_t \omega+(\mathbf u \cdot \nabla_{\mathbf r}) \omega=\nu \nabla^2 \omega+\nabla \times \mathbf F 
    \label{eq:Vor}
\end{equation}
(which can be derived by taking the curl of \eqref{eq:NS}).
We shall also analyze how a magnetic field $B \hat{\mathbf z}$ affects flows. There are several advantages to this approach. It is suggested in \cite{FGS22} that a circular geometry might be more appropriate for turbulent flows. Furthermore, we have the added benefit that vorticity is easier to solve for in this geometry. Also, the $z$ component of the magnetic field due to the motion of electrons (which could be experimentally observed) depends more naturally on the vorticity of the electron fluids up to a boundary contribution \nocite{Ja21} \footnote{This expression can be derived using standard identities \protect\cite{Ja21} in vector calculus and the Biot--Savart Law.}

\begin{equation}
    B_{z,e}(\mathbf r)=-\frac{ne\mu_0}{4\pi}\left(\oint_{\partial \mathcal S}\frac{\mathbf u(\mathbf r')\times d\mathbf r'}{|\mathbf r-\mathbf r'|} +\int_{\mathcal S}d^2\mathbf r'\frac{\omega(\mathbf r')}{|\mathbf r-\mathbf r'|}\right).
\label{eq:Bom}
\end{equation}
Finally, we can determine the velocity field as well from the vorticity and the incompressibility criterion $\nabla\cdot \mathbf u=0$,

\begin{equation}
    \mathbf u(\mathbf r)=\oint_{\partial \mathcal S} \frac{d\mathbf r'\cdot \mathbf u(\mathbf r')}{2\pi}\frac{\mathbf r-\mathbf r'}{|\mathbf r-\mathbf r'|^2}- \int_{\mathcal S}\frac{ d^2\mathbf r' \omega(\mathbf r') \hat{\mathbf z}}{2\pi}\times\frac{\mathbf r-\mathbf r'}{|\mathbf r-\mathbf r'|^2}.
    \label{eq:uom}
\end{equation}

In the next two subsections, we study the system in the presence and absence of a magnetic field.

\subsection{In the absence of a Magnetic Field}
\label{ssc:nmf}

We shall first consider the case when there is no external magnetic field. We linearize \eqref{eq:Vor} around the equilibrium solution $\mathbf u=\frac{\gamma}{r}\mathbf{\hat r}\equiv \bar u_r \mathbf{\hat r}$ to arrive at
\begin{equation}
     \partial_t \omega + \bar u_r \partial_r \omega=\nu \nabla^2 \omega .
     \label{eq:our}
\end{equation}
We follow \cite{LL987} and Fourier decompose $\omega$ with respect to time and azimuthal angle $\theta$ as $\omega=\Re(\sum w(n,\Omega) e^{-ij\theta-i\Omega t})$. Upon using this substitution, our differential equation reduces to
\begin{equation}
    r^2 \partial_r^2w+r\frac{\nu-\gamma}{\nu}\partial_r w -j^2w+\frac{i\Omega}{\nu} r^2w=0.
    \label{eq:wir}
\end{equation}
This can be solved exactly using Bessel functions \cite{AS948}

\begin{equation}
    w(j, \Omega,r)=r^{\frac{\gamma}{2\nu}}\left(\mathcal A(j,\Omega) I_{\sqrt{4\nu^2 j^2+\gamma^2}/(2\nu)}\left(e^{3\pi i/4}\sqrt{\frac{\Omega}{\nu}}r\right) + \mathcal B(j,\Omega)K_{\sqrt{4\nu^2 j^2+\gamma^2}/(2\nu)} \left(e^{3\pi i/4}\sqrt{\frac{\Omega}{\nu}}r\right)  \right),
    \label{eq:wrO}
\end{equation}
where $I$ and $K$ are the modified Bessel functions.
\subsection{In the presence of a Magnetic Field}
\label{ssc:ymf}

Let us now consider the case when there is an external magnetic field $B\mathbf {\hat z}$. Our analysis proceeds similarly as the previous case. We linearize around the solution $\mathbf u=\frac {eB}{m}r \bm{\hat  {\theta}}\equiv \bar u_\theta \bm{\hat \theta}$ to arrive at
\begin{equation}
    \partial _t \omega+\frac{u_{\theta}}{r} \partial _{\theta}\omega=\nu \nabla ^2\omega.
        \label{eq:wut}
\end{equation}
Once again doing a Fourier decomposition, we arrive at
\begin{equation}
    r^2\partial_r^2 w+r\partial_r w  - j^2 w+\frac{ir^2}{\nu}\left(\Omega+\frac{eBj}{m}\right)w=0,
    \label{eq:wit}
\end{equation}
which can also be solved using Bessel functions as 
\begin{equation}
    w(j , \Omega, r)=\mathcal C(j,\Omega)I_{j}\left(e^{3\pi i/4}\sqrt{\frac{\Omega}{\nu}+\frac{eBj}{m\nu}}r\right)+\mathcal D(j, \Omega)K_j\left(e^{3\pi i/4}\sqrt{\frac{\Omega}{\nu}+\frac{eBj}{m\nu}}r\right).
    \label{eq:wBO}
\end{equation}
\subsection{Qualitative Behavior and Asymptotics}
\label{ssc:qba}

Due to the contribution of the modified Bessel function of the first kind $I$, the asymptotic of $w$ (up to a constant factor) as $r\rightarrow \infty $ \cite{AS948}

\begin{equation}
    w_{B= 0} \sim r^{\gamma/(2\nu)-1/2}\exp\left(\sqrt{  {\frac{\Omega}{\nu}}}\left(\frac{r}{\sqrt{2}}-i\frac{r}{\sqrt 2}\right)\right),
    \label{eq:wrg}
\end{equation}
\begin{equation}
    w_{B,j} \sim r^{-1/2}\exp\left(\sqrt{  {\frac{\Omega}{\nu}+\frac{eBj}{m\nu}}}\left(\frac{r}{\sqrt{2}}-i\frac{r}{\sqrt 2}\right)\right).
    \label{eq:wrB}
\end{equation}
(This result, however, does not hold if the coefficient of $I$ is $0$ as the modified Bessel function of the second kind $K$ goes to $0$ for large $r$.) Observe that $\Re(we^{-in\theta})$ varies a lot with $\theta$ when $r$ is large. Hence, strong changes in the vorticity can be expected near the boundary of our geometry (until boundary effects dominate). This behavior should be observable in the velocity and magnetic fields as well, as shown in the figures in the next section.

\subsection{Vorticity and Velocity}
\label{ssc:vmf}
In this subsection, we present the graphical representations of our calculations. We assume that the effective mass of the electrons to be $m=0.03 m_e=2.7\times 10^{-32}\ \mathrm{kg}$  (applicable to a bi-layer graphene),\cite{TTGP15}  the viscosity $\nu =0.1\ \mathrm{m^2s^{-1}}\ $ \cite{BXP19} and a Corbino Geometry of inner and outer radii $r_{\mathrm{min}}=2\times 10^{-6}\ \mathrm{m}$ and $r_{\mathrm{max}}=6\times 10^{-6} \mathrm{m}$. \cite{KBS22} When $B$ is $0$, we assume the velocity associated with the stable flow to be $u\sim 100\ \mathrm{ms^{-1}}$, \cite{TTGP15} hence we take $\gamma=4\times 10^{-4}\ \mathrm{m^2s^{-1}}$. When $B$ is not zero, we take $B=0.01\ \mathrm {T}$ .\cite{TVP14} We assume the order of magnitude of $\Omega$ ranges from high megahertz to low terahertz. \cite{TVP14,FGS22} 

\begin{figure}[h]
  \begin{subfigure}{\linewidth}
  \includegraphics[width=.3\linewidth]{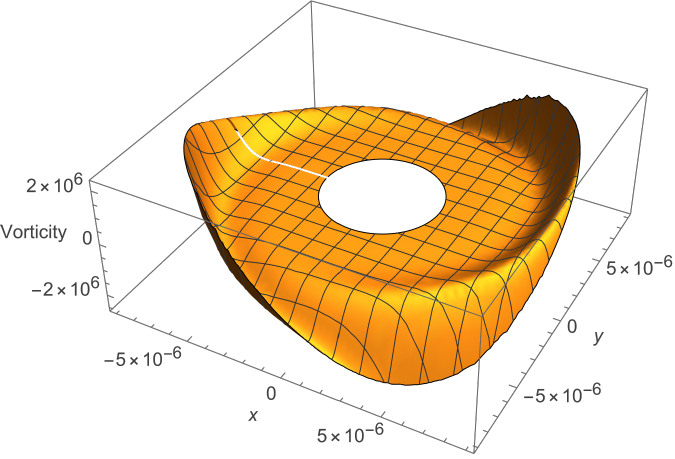}
  \includegraphics[width=.3\linewidth]{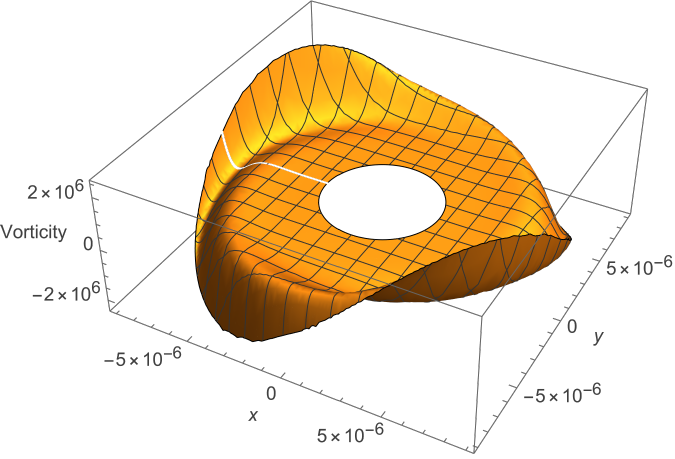}
  \caption{Vorticity (in SI units) when $B=0.00\ \mathrm T$ at $t=0\ \mathrm{s}$ (left) and $t=3\pi\times 10^{-12}\ \mathrm{s}$ (right)}
  \end{subfigure}
  \begin{subfigure}{\linewidth}
  \includegraphics[width=.3\linewidth]{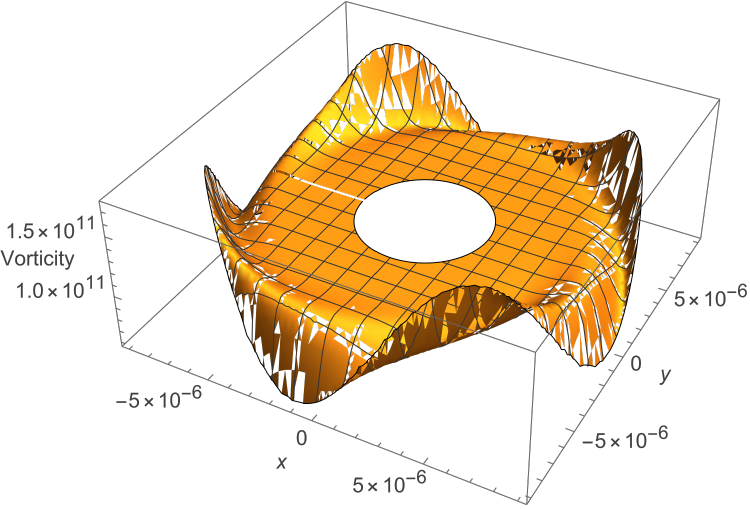}
  \includegraphics[width=.3\linewidth]{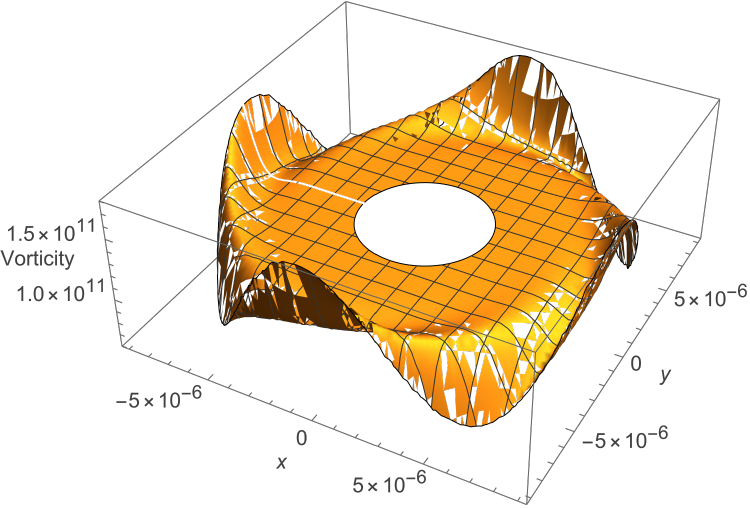}
  \caption{Vorticity (in SI units) when $B=0.01\ \mathrm T$ at $t=0\ \mathrm{s}$ (left) and $t=3\pi\times 10^{-12}\ \mathrm{s}$ (right)}
  \end{subfigure}
  \caption{Vorticity field in a Corbino geometry with radii $r_{\mathrm{min}}=2\times 10^{-6}\ \mathrm{m}$ and $r_{\mathrm{max}}=6\times 10^{-6} \mathrm{m}$ }
  \label{fig:Vor}
\end{figure}

\begin{figure}[h]
  \begin{subfigure}{\linewidth}
  \includegraphics[width=.3\linewidth]{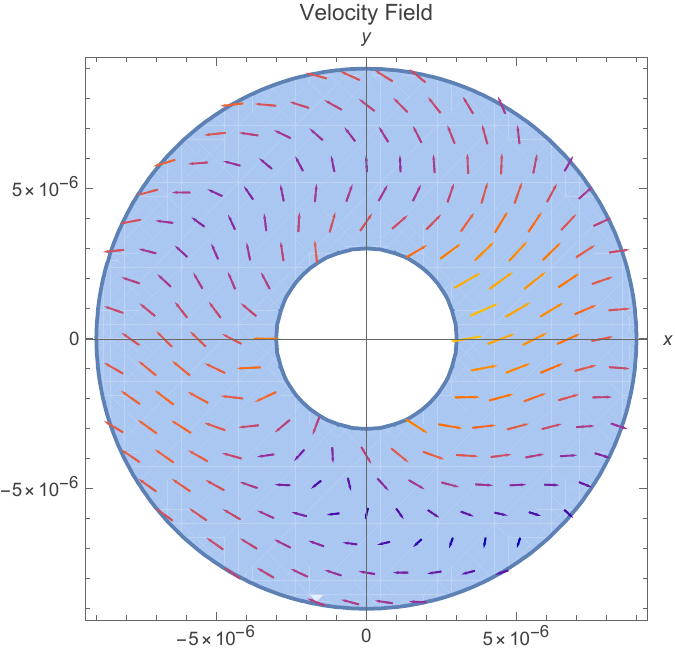}
  \includegraphics[width=.3\linewidth]{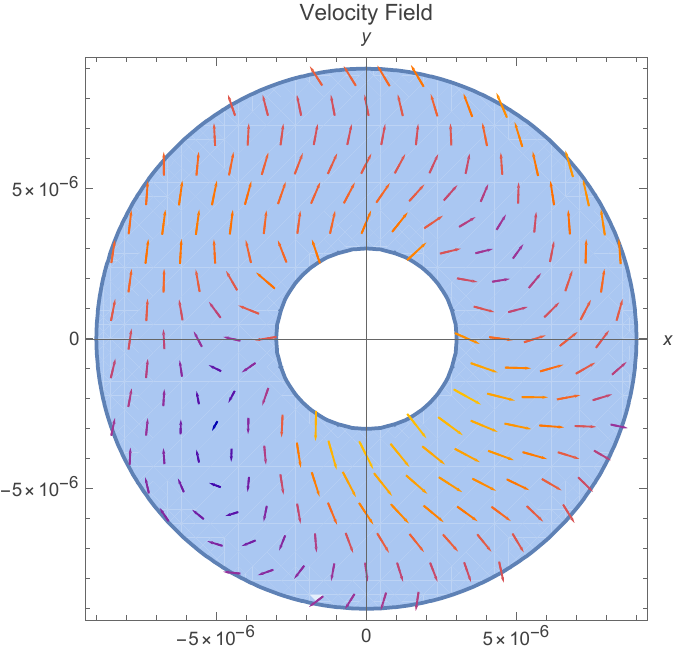}
  \caption{Velocity field when $B=0.00\ \mathrm T$ at $t=0\ \mathrm{s}$ (left) and $t=3\pi\times 10^{-12}\ \mathrm{s}$ (right)}
  \end{subfigure}
  \begin{subfigure}{\linewidth}
  \includegraphics[width=.3\linewidth]{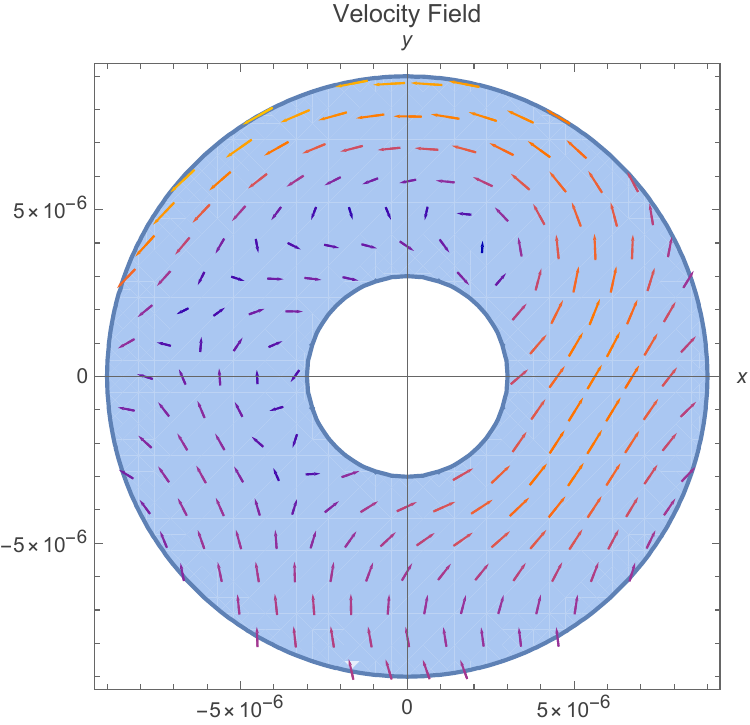}
  \includegraphics[width=.3\linewidth]{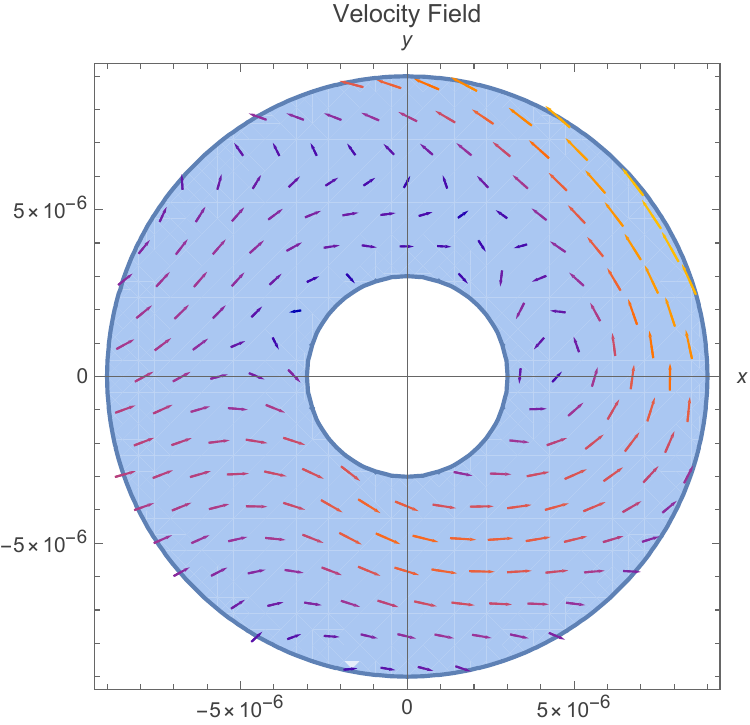}
  \caption{Velocity field when $B=0.01\ \mathrm T$ at $t=0\ \mathrm{s}$ (left) and $t=3\pi\times 10^{-12}$ (right)}
  \end{subfigure}
  \caption{Velocity field in a Corbino geometry with radii $r_{\mathrm{min}}=2\times 10^{-6}\ \mathrm{m}$ and $r_{\mathrm{max}}=6\times 10^{-6} \mathrm{m}$. The length of arrows represents the strength of velocity field. (The color serves as an aid for our eyes, where yellow indicates a stronger velocity while blue indicates a weaker velocity.)}
  \label{fig:Vel}
\end{figure}

The relevant values for the coefficients in \eqref{eq:wrO} and \eqref{eq:wBO} are given in Tables \ref{tab:B0} and \ref{tab:Bn}. The computations have been been evaluated at times $t=0\ \mathrm{s}$ and $t=3\pi \times 10^{-12}\ \mathrm{s}$.  In order to determine the velocity field, we need to introduce coefficients that shall serve the purpose of boundary conditions. We do so by writing $\mathbf u(\mathbf r)=\Re\left(\sum e^{-ij\theta-i\Omega t}(u_r(r)\mathbf{\hat r}+u_{\theta}(r)\mathbf{\hat \theta})\right)$ and providing the value of $u_r(r)$ at $r_{\mathrm{min}}$ and $r_{\mathrm{max}}$ as given in \ref{tab:B0} and \ref{tab:Bn}. The vorticity and velocity fields have been computed using \textit{Mathematica} and displayed in \ref{fig:Vor} and \ref{fig:Vel}. The length of the arrows in the velocity field diagrams is proportional to the strength of the velocity. (The color serves as an aid for our eyes, where yellow indicates a stronger velocity while blue indicates a weaker velocity.)

In this work, we do not compute the magnetic fields due to the motion of electrons. Interestingly, under some suitable boundary conditions, we can approximate $B_{z,e}\sim -ne\mu_0 R\omega$ where $R$ is a characteristic length of our geometry. \cite{TTGP15} Hence, the vorticity field can be seen as a crude approximation of the magnetic field scaled by some constant.

\begin{table}[h]
    \centering

        \caption{Coefficients of vorticity in \eqref{eq:wrO} and velocity boundary conditions of modes (in SI units) when $B = 0.00\ \mathrm T$}
        \label{tab:B0}
        \begin{tabularx}{\linewidth}{X X X X X X} 
            \hline\hline
            $j$ &$\Omega$& $\mathcal A$ & $\mathcal B$& $u(r_{\mathrm{min}})$ & $u(r_{\mathrm{max}})$ \\
            \hline
            $1$ & $1\times 10^{11}$ & $1\times 10$ & $(i+2)\times 10$& $2\times 10$ & $3i\times 10$ \\
            \hline
            $2$ & $-4\times 10^{11}$& $(1+2i)\times 10$ & $2\times 10$& $(i+3)\times 10$ & $4\times 10 $ \\
            \hline
            $4$ & $3\times 10^{11}$ & $3i\times 10$ & $2\times 10$&$(2i+1)\times 10$& $1\times 10$ \\
            \hline\hline
        \end{tabularx}

    \end{table}

    \begin{table}[h]
   \centering
    
        \caption{Coefficients of vorticity in \eqref{eq:wBO} and velocity boundary conditions of modes (in SI units) when $B = 0.01\ \mathrm T$} 
        \label{tab:Bn}
         \begin{tabularx}{\linewidth}{X X X X X X} 
            \hline\hline
            $j$ &$\Omega$& $\mathcal C$ & $\mathcal D$& $u(r_{\mathrm{min}})$ & $u(r_{\mathrm{max}})$ \\
            \hline
            $1$ & $1\times 10^{11}$ & $1\times 10^{5}$ & $(i+2)\times 10^{5}$& $2\times 10^{5}$ & $3i\times 10^{5}$ \\
            \hline
            $2$ & $-4\times 10^{11}$& $(1+2i)\times 10^{5}$ & $2\times 10^{5}$& $(i+3)\times 10^{5}$ & $4\times 10^{5} $ \\
            \hline
            $4$ & $3\times 10^{11}$ & $3i\times 10^{5}$ & $2\times 10^{5}$&$(2i+1)\times 10^{5}$& $1\times 10^{5}$ \\
            \hline\hline
        \end{tabularx}
    
    \end{table}

\section{Conclusions and Future Work}
\label{sec:cfw}

In this work, we have obtained a correction of the form $W^4$ to the conductivity and analyzed the frequency of perturbations. We have also examined various aspects of the Corbino geometry in a magnetic field and studied how vorticity strongly varies at the boundary of the geometry. However, our approaches have limitations that could be advanced in later works. The treatment ignores the effect of spin as well as damping forces that are proportional to the velocity. Also, there might be better, more rigorous ways of using Kraichnan's theory in making calculations in electron hydrodynamics. Furthermore, more numerical methods are required to better understand turbulence in the spherically symmetric geometries. 

\section{Acknowledgments}

I would like to thank Professor Subroto Mukerjee (Indian Institute of Science), Professor Arindam Ghosh (Indian Institute of Science) and Professor Samriddhi Sankar Ray (International Center for Theoretical Sciences) for their valuable guidance, advice and support.

\bibliographystyle{unsrtnat}
\bibliography{biblio}

\end{document}